\documentclass[showpacs,aps,prl,twocolumn]{revtex4}
\usepackage{graphicx}% Include figure files
\usepackage{dcolumn}% Align table columns on decimal point
\usepackage{hyperref}% add hypertext capabilities
\usepackage{psfrag,bm,amsmath,amsfonts,amssymb,color}
\usepackage{ulem}
\usepackage{times}
\usepackage{mathrsfs}
\begin{document}

\def\sss{\scriptscriptstyle}

\def\bhm{M_{\bullet}}
\def\bhl{L_{\bullet}}
\def\dotm{\dot{m}}
\def\dotM{\dot{M}_{\bullet}}
\def\dotMedd{\dot{M}_{\rm Edd}}
\def\dv{\Delta v}
\def\ergs{\rm erg~s^{-1}}
\def\fblr{f_{\sss\rm BLR}}
\def\feii{Fe {\sc ii}}
\def\fx{f_{\sss{\rm X}}}
\def\fbh{f_{\bullet}}
\def\kms{\rm km~s^{-1}}
\def\kappabol{\kappa_{_{\rm B}}}
\def\ledd{L_{\rm Edd}}
\def\mbh{M_{\bullet}}
\def\mubh{\mu_{\bullet}}
\def\msun{M_{\odot}}
\def\mathc{\mathscr{C}}
\def\oiii{[O\,{\sc iii}]}
\def\qdotm{q_{_{\dotm}}}
\def\rblr{R_{\sss{\rm BLR}}}
\def\sunm{M_\odot}
\def\vfwhm{V_{\sss{\rm FWHM}}}
\def\vhbeta{V_{_{\rm H\beta}}}
\def\vmgii{V_{_{\rm Mg~II}}}
\def\Dbh{{\cal D}_{\bullet}}
\def\sigmaT{\sigma_{\sss{\rm T}}}
\def\taut{\tau_{\sss{\rm T}}}

\title{Super-Eddington accreting massive black holes as long-lived cosmological standards}

\author{Jian-Min Wang$^{1,2}$}\email{wangjm@ihep.ac.cn}
\author{Pu Du$^1$}
\author{David Valls-Gabaud$^{3,1,2}$}
\author{Chen Hu$^1$}
\author{Hagai Netzer$^4$}

\affiliation{%\footnotesize
$^1$Key Laboratory for Particle Astrophysics, Institute of High Energy Physics, CAS, 19B Yuquan Road,
Beijing 100049, China}
\affiliation{%\footnotesize
$^2$National Astronomical Observatories of China, CAS, 20A Datun Road, Beijing 100020, China}
\affiliation{%\footnotesize
$^3$LERMA, CNRS UMR 8112, Observatoire de Paris, 61 Avenue de l'Observatoire, 75014 Paris, France}
\affiliation{%\footnotesize
$^4$School of Physics and Astronomy and The Wise Observatory, The Raymond and Beverley Sackler
Faculty of Exact Sciences, Tel-Aviv University, Tel-Aviv 69978, Israel}

\date{Received 27 August 2012; accepted 16 January 2013 by Physical Review Letters}
%\date{\today}

\begin{abstract}
Super-Eddington accreting massive black holes (SEAMBHs) reach saturated luminosities above a certain
accretion rate due to photon trapping and advection in slim accretion disks. We show that these
SEAMBHs could provide a new tool for estimating cosmological distances if they are properly identified
by hard X-ray observations, in particular by the slope of their 2--10 keV continuum.
To verify this idea we obtained black hole mass estimates and X-ray data for a sample of
60 narrow line Seyfert 1 galaxies that we consider to be the most promising SEAMBH candidates.
We demonstrate that the distances derived by the new method for the objects in the sample
get closer to the standard luminosity distances as the hard X-ray continuum gets steeper.
The results allow us to analyze the requirements for using the method in future samples of active
black holes and to demonstrate that the expected uncertainty, given large enough samples, can make
them into a useful, new cosmological ruler.
\end{abstract}
\pacs{98.80.Es, 98.54.Cm, 98.62.Js, 98.62.Mw, 95.36.+x}

\maketitle
The discovery of the accelerating expansion of the Universe has now been established through observations
of type Ia supernovae (SNe Ia) \cite{snia}, and is likely to be confirmed further with new standard rulers
provided by baryon acoustic oscillations, weak lensing and clusters of galaxies \cite{frieman08}. However,
SNe Ia beyond $z\gtrsim 1.5$ are rare \cite{rubin12} as there is no time for their progenitors to evolve
in substantial numbers given their lower metallicity \cite{kobayashi09}. To further probe the dynamics
of the acceleration, new distance indicators are needed at and beyond these redshifts. Based on well-understood
physics we show  in this Letter that super-Eddington accreting massive black holes (hereafter SEAMBHs)
in some active galactic nuclei (AGNs), that are characterized by a  mass of $10^{6\sim8}\sunm$,
can provide a new tool to estimate cosmological distances at a wide range of redshifts, including the high
redshift Universe.

Radiation pressure limits the spherical accretion rate onto black holes to
$\dot{M}_{\rm Edd}=L_{\rm Edd}/\eta c^2$, where $L_{\rm Edd}=4\pi G\bhm m_{\rm p}c/\sigmaT$ is the
Eddington luminosity for a pure hydrogen plasma, $\eta$ ($ \sim 0.1$) is the mass to radiation
conversion efficiency, $\sigmaT$ is the Thomson cross section, $m_{\rm p}$ is the proton mass, $c$
is the speed of light, $G$ is the gravitational constant and $\bhm$ is the
black hole  mass. However,
super-Eddington accretion onto black holes
is feasible in slim disks where the radiation
pressure-dominated regions (RPDR) are thermally stable due to the radial advection of the locally
emitted radiation \cite{abramowicz88}. In such disks, the timescale of photon diffusion to the disk
surface is longer than that of the radial motion of the accreting gas in the RPDR. The photons are
trapped inside the accretion flows and are advected into the black holes. This advection
dominates within the photon trapping radius, $R_{\rm trap}\approx 430~\dotm_{15} R_{\rm g}$,
where $\dotm_{15}=\dotm/15$, $\dotm=\dotM/\dot{M}_{\rm Edd}$, $\dotM$ is the mass accretion rate
and $R_{\rm g}=G\mbh/c^2$ \cite{slim_adaf}. Photon trapping affects the total emitted radiation and
results in a saturated luminosity, $\bhl$, which is proportional (logarithmically, rather than linearly)
to the accretion rate \cite{slim_adaf,slim_sed},
\begin{equation}
\bhl=\ell_0\left(1+a\ln \dotm_{15}\right)\mbh \; ,
\label{eqn:1}
\end{equation}
where $\ell_0\approx 5.29\times 10^{38}~~{\ergs}~M_{\odot}^{-1}$, and $a\approx 0.476$ \cite{slim_sed}.
For reference, at $\dotm=15$ the saturated luminosity is  $\sim 4.20L_{\rm Edd}$. Thus, at a given
black hole mass, SEAMBHs are radiating basically at a constant luminosity which, as shown below, can
therefore be used to deduce cosmological
distances.

In this Letter we address two important issues: how to identify SEAMBHs and how to test, observationally,
Equation~\ref{eqn:1} and its uncertainties such that it can be used to derive reliable cosmological distances.
While SEAMBHs are predicted to have unique optical-UV spectral
characteristics \cite{abramowicz88,slim_adaf,slim_sed}, their use to identify such sources is hampered by
the dilution of the disk emission by stellar radiation from the host galaxy at long wavelengths,
and by the Galactic and inter-galactic absorption at short wavelengths. In fact, current
observations cannot identify such systems using only their spectra in the optical-UV
 domain. Fortunately,
X-ray spectroscopy  allows such identifications for two reasons. First, there is a well-known positive
correlation between the 2-10 keV X-ray photon spectral index ($\Gamma$) and the Eddington ratio
($L_{\rm Bol}/L_{\rm Edd}$) \cite{Gamma2-10}. In addition, higher
$L_{\rm Bol}/L_{\rm Edd}$ sources emit a smaller fraction of their total radiation ($\fx=L_X/\bhl$) at hard
X-ray energies \cite{factor_X}. These properties are easy to measure with modern X-ray observations and are
similar to those observed in Galactic black holes in
X-ray binaries \cite{remillard06}.

The general theory that links X-ray emission to the optical-UV spectrum of accretion disks is based on
the assumption of a hot corona above the disk that is the source of the X-ray radiation. The X-ray
emission efficiency of the processes in the corona depends on $L_{\rm Bol}/L_{\rm Edd}$. In particular,
the magneto-rotational instability is a key factor to produce the viscosity that helps
transporting angular momentum outward \cite{balbus98}. The process takes place through magnetic buoyant
transportation above the cold disk \cite{galeev79} which leads to a ``corona-dominated" dissipation through
hard X-ray emission. The increases of the accretion rates result in the weakening of the transportation of
magnetic tubes  because of the inflation of the disk by radiation pressure which yields a
reduction of the buoyant velocity. The end result of these processes is the radial advection of the emitted
photons and the suppression of the relative X-ray flux ($\fx$) of the system \cite{merloni03}.

The second effect of an increasing $L_{\rm Bol}/L_{\rm Edd}$ ratio is the steepening of the X-ray photon index
$\Gamma$. There have been various attempts at calculating $\Gamma$ in hot coronae from basic considerations
of the conditions in accretion disks \cite{galeev79}. The 2-10 keV emission is mainly due to the
Comptonization of photons from the cold disks (the X-ray reflection can be neglected in this
band \cite{x-ray-reflection}). It has been shown that $\fx\propto \mbh^{-1/18}\dotm^{-4/9}$ in the RPDR (see
equation 13 for $\fx\ll 1$ and Figure 1 in Ref. \cite{merloni03}) and thus one expects that a luminosity
$\fx\bhl$ is radiated by the coronae. Since  Comptonization is the main cooling process, the balance
between heating and cooling yields the density of hot electrons $n_c\propto \fx$ and
$n_c\propto \dotm^{-4/9}$.

Under the conditions discussed above, the X-ray photon index can be approximated
by $\Gamma\approx 2.25y^{-2/9}$ \cite{beloborodov99}, where $y=4\theta_e\taut(1+4\theta_e)(1+\taut)$ is
the Comptonization parameter, $\theta_e=kT_e/m_ec^2$, $T_e$  is the electron temperature
and $\taut$ is the Thomson scattering optical depth. It is expected that
$y\propto n_c^{\gamma}\propto \dotm^{-4\gamma/9}$, where $\gamma=1$ for unsaturated Comptonization
($\taut<1$) and $\gamma=2$ for saturated Comptonization. This scenario is supported by the
behaviors of black hole
X-ray binaries in very high states \cite{remillard06}, showing that the 2-10 keV
emission is at the level of low/hard states, but the photon indexes are typically $\Gamma>2$.
Considering that hard X-ray spectra have a cutoff of 50-100 keV ($\theta_e\sim 0.1-0.2$) \cite{cutoff}
and applying the above coronal model to AGNs, we have $\Gamma\propto \dotm^{8\gamma/81}$.
This  prediction agrees with the observed $\Gamma-L_{\rm Bol}/L_{\rm Edd}$ correlation of
AGNs hosting standard (optically thick and geometrically thin) accretion disks \cite{Gamma2-10}.

The simplest model for the slim disks of SEAMBHs assumes  a spherical hot corona with
a characteristic size $\ell_c$. The Comptonization of photons from the slim disk surface produces a
hard X-ray luminosity
$L_X=4\theta_en_e\sigmaT c\left(\bhl/4\pi \ell_c^2c\right)\left(4\pi\ell_c^3/3\right)$, giving rise to
$\taut=3\fx/4\theta_e\approx 0.8~f_{0.1}\theta_{0.1}^{-1}$, where $\taut=n_e\sigmaT\ell_c$,
$f_{0.1}=\fx/0.1$ and $\theta_{0.1}=\theta_e/0.1$. The coronae having a Comptonization parameter of
$y\approx 0.8$, the SEAMBHs are then characterized by $\Gamma\gtrsim 2.3y_{0.8}^{-2/9}$, where
$y_{\sss 0.8}=y/0.8$. Obviously there are uncertainties in these parameters and the resulting theoretical
relationships, however, we can use the observed $\Gamma$ to identify SEAMBHs in different types of
 active galactic nuclei.

The best group of AGNs where such processes have been studied are narrow line Seyfert 1
galaxies (NLS1s). These objects are separated from broad line Seyfert 1 galaxies (BLS1s)
by having the following properties \cite{osterbrock85,boller96}:
1) the full-width-half-maximum (FWHM) of H$\beta$ profiles are narrower than $2000 \, \kms$;
2) strong soft X-ray excess;
3) unusually strong (relative to H$\beta$) optical iron emission lines;
4) weak \oiii\ lines (\oiii/H$\beta<3$);
5) fast, large amplitude X-ray variations \cite{leighly99}.
The typical black hole mass in NLS1s (see below) is considerably smaller than  that
in BLS1s of similar $L_{\rm Bol}$
which implies that many of them may be accreting at super-Eddington rates.

To test observationally  Equation~\ref{eqn:1} in NLS1s, we have to estimate $\mbh$.  Black hole
masses can be  measured individually using the reverberation mapping (RM) technique, which invokes the
response (time lag)  of the broad emission lines with respect to changes in the continuum produced
by the underlying disk  \cite{kaspi00,pancoast11}. In its most detailed version, the
velocity-resolved reverberation mapping (VRRM), one can derive the spatial distribution of the line
emitting gas, and its velocity, at every location around the black hole.
In principle, this phase-space
mapping  enables us to determine, accurately, the black hole mass (this is equivalent to dynamical methods
which are used to measure the masses of black holes in normal nearby galaxies). Such two dimensional mappings
are only available for a handful of sub-Eddington sources and many details of the technique need to be improved
for accurate measurements (e.g. the case of the BLS1 Mrk 50 \cite{pancoast11}).

Fortunately, we can use as a proxy the tight correlation between the size of the broad line regions (BLR)
(the time lag times the speed of light) and the underlying continuum luminosity to obtain an empirical
relationship, for a sample of about 35 AGNs, that can be combined with the observed line
widths to estimate
the black hole masses in a large sample of sources. The relationship is given by
\begin{equation}
\rblr=R_0 \left ( \frac{L_{5100}}{10^{44}~{\rm erg~s^{-1}}} \right )^{\alpha} \; ,
\label{eqn:2}
\end{equation}
where $R_0 \simeq 9\times 10^{16}$cm, $\alpha=0.6\pm 0.1$, and $L_{5100}$ is the AGN continuum luminosity
($\lambda L_{\lambda}$ at 5100 \AA\, in units of $10^{44} \ergs$) \cite{kaspi00,bentz09}. We note that among
the 35 AGNs used to derive this correlations, eight are NLS1s and follow the same trend as other
sources \cite{bentz11}.

\begin{figure*}%[t!]
\begin{center}
\includegraphics[angle=-90,width=0.98\textwidth]{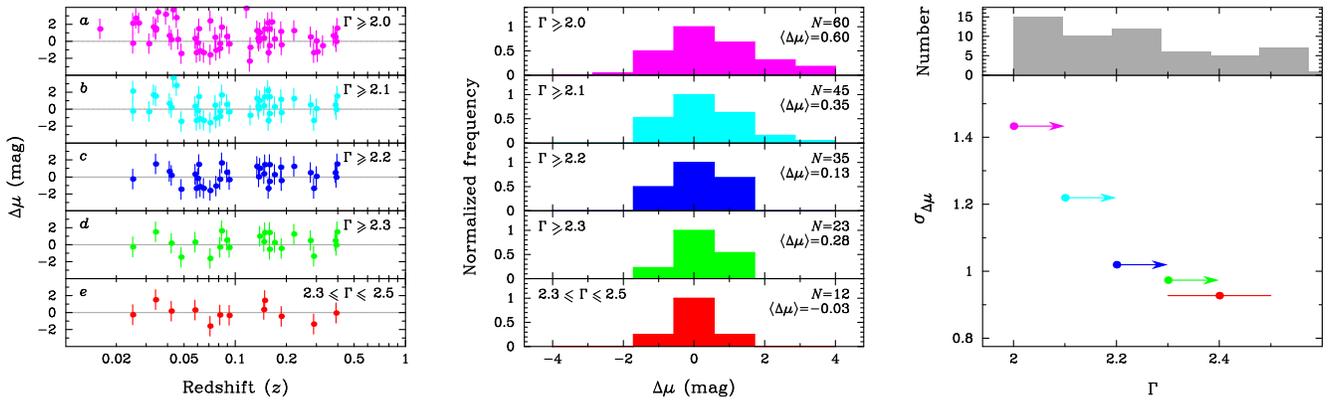}\end{center}
\caption{\footnotesize {\it Left}: Residuals of the SEAMBH distance modulus ($\Delta\mu=\mubh-\mu_{\rm L}$).
The error bars are taken as $\Delta\mubh=1.17$ mag (see text for details). The assumed cosmological parameters
are $H_0=71~{\rm km~s^{-1}~Mpc^{-1}}$, $\Omega_{\Lambda}=0.73$ and $\Omega_M=0.27$. {\it Middle}: The distribution
of $\Delta\mu$ as a function of the  hard X-ray (2-10 keV) photon index $\Gamma$. The normalized frequency is
referred to fractions to the peak number of objects. $N$ is the total number of  SEAMBHs selected by $\Gamma$.
{\it Right}: scatter of the $\Delta\mu$ distributions with $\Gamma$ (bottom) and the distribution of $\Gamma$
for the sample of SEAMBH candidates. The dependence of the dispersion in the residuals as a function of
$\Gamma$ shows a systematic decrease and tends to $\sigma_{\Delta\mu}^{\rm min}$, indicating the efficacy of
the selection as standard candles.}
\label{fig:1}
\end{figure*}

Equation~\ref{eqn:2} enables  us to obtain the black hole masses by assuming a virialized (gravitationally bound)
cloud system and combining all the unknown geometrical factors, such as the inclination to the line
of sight, into a single constant $\fblr$. Using this constant we can now write an expression for the
``reverberation mapping-based virial mass estimate" of super-massive black holes as
\begin{equation}
\bhm=\fblr G^{-1}\vfwhm^2\rblr \; ,
\end{equation}
where $\vfwhm$ is the FWHM of the broad emission line (e.g. H$\beta$) that was used to derive the time
lag in the RM measurement. The factor $\fblr$ is calibrated by comparing the results of the RM experiments
to direct black hole measurements based on the  $\bhm-\sigma_*$ relation, where $\sigma_*$ is the stellar
velocity dispersion in the bulge of the host galaxy. Such a comparison is now available for about 30 out
of the 35 AGNs in the RM sample. It shows that $\fblr \simeq 1.2 \pm0.2$ \cite{onken04}. Noting that
$\bhl= \kappabol L_{5100}$, where $\kappabol$ is a bolometric correction factor, we obtain an expression
of $\bhl$ in terms of $\fblr$ and $\vfwhm$. Since $L_{5100}= 4 \pi \Dbh ^2 F_{5100}$, where $F_{5100}$ is
the measured  continuum flux in units of ${\rm erg~s^{-1}~cm^{-2}}$ at 5100 \AA\
and $\Dbh$ is the luminosity distance of the black hole, we obtain the  expression
\begin{equation}
\Dbh=\frac{1}{\sqrt{4\pi}} \left[ \frac{ l_0\left(1+a\ln \dotm_{15}\right)\fblr R_0 }{G\kappabol} \right]^{1/2(1-\alpha)}
      \frac{ \vfwhm^{1/ (1-\alpha)}}{ F_{5100}^{1/2}}.
\label{eqn:4}
\end{equation}
This expression still involves the unknown accretion rate $\dotm_{15}$ which, as suggested earlier,
can be estimated from the X-ray slope $\Gamma$. However, the dependence for $\Gamma > 2$ (which is the
one we are interested here, see below) is weak enough that we can use the approximation
$\dotm_{15}=1$ and obtain $\Dbh$ from the other known observables and constants.
Thus, having established that an object is a SEAMBH by measuring its $\Gamma$ index, we have a way
to directly measure its distance.

In the following we use the distance modulus, $\mubh=5\log\left(\Dbh/{\rm pc}\right)-5$, and compare
it with the one obtained from the standard luminosity distance ${\cal D}_{\rm L}$ in the
Friedman-Lema\^{\i}tre-Robertson-Walker metric $\mu_{\rm L}=5\log \left({\cal D}_{\rm L}/{\rm pc}\right)-5$.
The prediction is that comparing $\Dbh$ and $D_{\rm L}$ we will get smaller residuals
$\Delta\mu=\mubh-\mu_{\rm L}$ for larger $\Gamma$, since large indices point to conditions closer to those
predicted by the SEAMBH theory.

We now turn to the available samples of SEAMBH candidates. While the observed NLS1 properties may all
be related to the large Eddington ratio \cite{pounds95,boller96}, not all NLS1s have super-Eddington accretion
rates. It is thus necessary to use the hard X-ray spectra to identify SEAMBHs among NLS1s. We selected
a large number of NLS1s from several heterogeneous samples \cite{boller96,samples} with hard X-ray
observations by {\it ASCA}, {\it XMM-Newton}, {\it Chandra} and {\it Swift} \cite{missions}. All data
and data reduction details are provided in the Supplementary Materials Online. In short, we use the
observed $F_{5100}$ flux and the estimated black hole mass $\mbh$ to derive $\kappabol$ and hence $\bhl$
for each source (see also a somewhat different approach but similar results in \cite{jin12,slim_sed}). The
contaminations by the  stellar light of the host galaxy were removed, prior to
the estimate of $F_{5100}$, using the approximation described in Materials Online. We examined the
$L_{\rm Bol}/L_{\rm Edd}$ distribution in our NLS1 sample and found that, indeed, many of them
indicate super-Eddington ratios, up to 5 and even more. Thus, the selection of sources by $\Gamma$ is
indeed a good way to identify such sources.

We calculate $\Delta\mu$ for all sources, bin them into various groups of different $\Gamma$, and
plot them in Figure~\ref{fig:1}. As shown by the standard deviation ($\sigma_{\Delta\mu}$), the
scatter of $\Delta\mu$ systematically decreases as $\Gamma$ increases.
From Equation~\ref{eqn:4}), this behavior
can be understood in terms of the scatter expressed by
\begin{equation}
\Delta\mubh=\frac{5}{\ln10}\left[\frac{(1-\alpha)^{-2}a^2}{4(1+a\ln\dotm_{15})^2}
          \left(\frac{\Delta\dotm}{\dotm}\right)^2+\Delta\mu_X^2\right]^{1/2},
\end{equation}
where $\Delta\mu_X^2=\sum_{i=1}^5\Delta\mu_i^2$,
$\Delta\mu_i=A_i\left(\Delta X_i/X_i\right)$, $X_i=\fblr$, $R_0$, $\kappabol$, $\vfwhm$, $F_{5100}$,
$A_i(i=1,2,3)=1/2(1-\alpha)$, $A_4=1/(1-\alpha)$ and $A_5=1/2$. $\Delta\mubh$
converges to $\Delta\mu_X$ as $\Delta\dotm/\dotm$ decreases with increasing $\Gamma$.

To illustrate the typical uncertainty on individual points we assume that
$\Delta \fblr /\fblr = 0.2$ \cite{onken04} (from the scatter in the $\mbh-\sigma_*$ distribution),
$\Delta F_{5100} / F_{5100} = 0.2 $ (from the known variations in the optical continuum and the uncertainty
in the substraction of the stellar background), $\Delta \vfwhm / \vfwhm = 0.05$ (from the uncertainty in
fitting the emission line profiles and measuring $\vfwhm$), $\Delta R_0 /R_0 = 0.2$ and
$\Delta \kappabol / \kappabol = 0.3$. The estimated uncertainty of $\kappabol$ is the largest and most
problematic for several reasons. First, all our estimates of $\kappabol$ are derived from theoretical
calculations of slim disk spectra \cite{slim_sed}. These have not been verified observationally because
of the lack of extreme UV observations, where most of the emitted
luminosity in disks around small black
holes ($10^6 - 10^7$ $\msun$) is emitted. Moreover, a factor of
10 increase in black hole mass results in a factor of $10^{1/3}$ decrease in $\kappabol$ (from about 100 in
$10^6 \msun$ to about 40 in $10^7 \msun$ black holes). The entire range of black hole masses in our sample
suggests a very large $\Delta \kappabol / \kappabol$. Fortunately, the individual masses are known and the
uncertainty on massive black holes  using the RM-based virial method is only a factor of about
3 \cite{vestergaard06}. This and the allowed range of spectral shapes and $\dot{m}$ gives
the quoted estimate on $\Delta \kappabol / \kappabol $.

Combining all these uncertainties and assuming $\alpha=0.6$ we obtain $\Delta\mu_X=0.54$.
This corresponds to $\Delta \mubh \simeq 1.17$ mag since the first term
in Equation~\ref{eqn:1} tends to
vanish for large enough values of $\Gamma$. This is not surprising given
Equation~\ref{eqn:1} and the known
uncertainties on $\mbh$ measured by the RM-based virial method. This uncertainty is estimated to be
0.3--0.5 dex which would suggest a similar uncertainty on $\bhl$. Since $\mu \propto 2.5 \log \bhl$,
we get a similar uncertainty on individually measured points to the one obtained by the more detailed
calculations. From Figure~\ref{fig:1}, we find $\Delta \mubh\gtrsim\sigma_{\Delta\mu}^{\rm min}=0.93$ mag, which
clearly shows that the selected sample has a small scatter of $\Delta\dotm/\dotm$ related to this term.

The panels of Figure~\ref{fig:1}  illustrate both the convergence to $\langle\Delta \mu \rangle= 0$ and
the reduced scatter when
using increasinglt larger values of $\Gamma$. The central panel shows that for a sufficiently steep
X-ray continuum, a combination of a large number of SEAMBHs gives, indeed, the correct distance
with a small scatter.

The systematic decrease of the dispersion in the subsamples with
increasing $\Gamma$,
while keeping a median with little variation, cannot be accounted by statistical fluctuations in the
subsamples. Considering that the subsamples are not independent, we assess the statistical
significance using Monte Carlo simulations with $10^7$ Gaussian samples of size $N=60$ with
the same mean and dispersion as the observed one. From them, we selected distinct (i.e.,
with no replacement) random sub-samples of the same size as the ones
selected (Figure~\ref{fig:1}),
and estimate the probabilities that the differences in the medians and the ratio of
the dispersions correspond to the observed ones.
We find that while the probability that the medians do not differ from the one of the
underlying sample of $N=60$ is always very large (above $\sim 80\%$), the probability
that the dispersions are as small as the observed one is always smaller than 3\%.
The observed trend cannot therefore be ascribed to random fluctuations of small
samples extracted from the main sample.
In the sample used here, there are only 12 sources
with $\Gamma\in[2.3,2.5]$ but future samples will be larger since such objects can be observed
to high redshift. We point out that the $\Gamma\ge2.3$ and
$2.3\le\Gamma\le 2.5$ panels are statistically indistinguishable because of the poor quality of the
sources with $\Gamma >2.5$. There are  11 sources in total with $\Gamma>2.5$ listed in Table 1 in the
Supplementary Material Online. Five of them with hard X-ray observations have large error bars
($\Delta\Gamma \ge 0.37$), making the $\Gamma-$binning less significant for these small samples.
Five other objects are observed in the 0.5--8 keV band by {\it Chandra} (Williams et al. 2004 in
\cite{samples}, see notes in Table 1), but the data quality only allows us to approximate the 0.5--8
keV spectral indexes with the 2--8 keV proxy, and the error bars of these sources remain very large
($\Delta \Gamma\gtrsim 0.3$ except one). The $2.3\le\Gamma\le2.5$ panel is shown for reference and
future observations. Moreover, the accuracy
of the measured $\mbh$ can increase substantially if new, dedicated RM experiments are carried out on a
large number of SEAMBH candidates. This can reduce the uncertainty on $\fblr$, $F_{5100}$, $\vfwhm$ and
$R_0$ and constrain $\alpha$ to be the slope for this population only. As Figure~\ref{fig:1} shows, it is reasonable
to assume that for a large enough sample of SEAMBHs at a narrow redshift range, we could expect a scatter
on the population $\mubh$ that approaches 0.15 mag, similar to the current accuracy of SNe Ia
method \cite{rubin12}.

SEAMBHs, as a new type of cosmological distance indicator, have a number of advantages over
others \cite{grijs11}:
(1) The saturated luminosities are well understood on physical grounds and have
no potential cosmic evolution.
(2) The objects can be efficiently selected by their X-ray or optical properties.
(3) They can probe a large range of redshifts, as they follow the cosmic growth of massive black holes
that are abundant at high-$z$ and are very luminous.
(4) Unlike SNe Ia, repeated observations can be made to improve the observational accuracy.

Several observational issues require careful attention: (1) As mentioned before, the $\rblr-L$ relation
applied here is the one obtained for {\it all} AGNs. The re-calibration of this relationship in a
dedicated NLS1s or SEAMBH sample, by  obtaining better
 estimates of $R_0$ and $\alpha$ in Equation~\ref{eqn:2},
 can reduce the scatter in the derived mass and hence $\kappabol$. Obviously, the estimated
$\mbh$ involves the distance to the source  which is the quantity we are attempting to measure.
However, we do not expect large differences between ${\cal D}_{\rm L}$ and $\Dbh$ so this
uncertainty, by itself, is very small. (2) The 2-10 keV luminosity and slope are both
variable \cite{leighly99} which may lead to misidentification of SEAMBHs. Long-term averaged values
can be used to improve the accuracy.

The prospects of building large samples of SEAMBHs to be used as tests of the current cosmological
model are promising. We expect that roughly 20\% of NLS1  with $\Gamma\ge 2$ host SEAMBHs. As NLS1
constitute about 10\% of all AGNs, there should be $10^4\sim 10^5$ SEAMBHs among  Seyfert 1 galaxies
with $z\le 0.3$ \cite{huchra92}. SEAMBHs could be even more abundant at high-$z$ although the definition
of NLS1 should be modified in such cases \cite{netzer07}. Here we require black hole mass estimates that
are based on both the  H$\beta$ and Mg {\sc ii} $\lambda$2798 \AA\ lines. X-ray spectra can be obtained
by {\it Nustar}, by the upcoming e{\it Rosita} and {\it HXMT} missions \cite{erosita}. Given accurate
observations of SEAMBHs at high redshift we will have a unique chance to explore in-depth the dynamics
of the accelerating Universe as well as the nature of dark energy.

\acknowledgments{
JMW thanks the hospitality of M. Ward and C. Done at Durham, where this work was initiated in early
October 2011. L. C. Ho, Y.-Y. Zhou, Z.-W. Han, C. Jin, Y.-R. Li, S.-M. Jia, J.-M. Bai and J.-C. Wang are
acknowledged for useful suggestions and discussions. The research is supported by 973 project (2009CB824800),
NSFC-11173023, -11233003, and -11133006. DVG was supported in part by the CAS with a Visiting Professorship
for senior international scientists.}

\def\apj{Astrophys. J.}
\def\apjs{Astrophy. J. Suppl.}
\def\aj{Astron. J.}
\def\aap{Astron. Astrophys.}
\def\mnras{Mon. Not. R. Astron. Soc.}
\def\araa{Ann. Rev. Astron. Astrophys.}
\def\pasj{Pub. Astron. Soc. Jap.}

\end{document}